**Two-stage short-term wind power forecasting algorithm using different feature-learning models**

Jiancheng Qin[a], Jin Yang[a], Ying Chen[b,*], Qiang Ye[b], Hua Li[c]

[a]School of Computer Science and Technology, Harbin Institute of Technology, Heilongjiang Province, China, 150000

[b]Department of Management Science and Engineering, School of Management, Harbin Institute of Technology, Heilongjiang Province, China, 150000

[c]Department of Mechanical and Industrial Engineering, Texas A&M University-Kingsville, Kingsville, Texas, U.S.A. 78363

*Corresponding author: Phone number: (+86)-451-86414019, Email: yingchen@hit.edu.cn

**Abstract**

Two-stage ensemble-based forecasting methods have been studied extensively in the wind power forecasting field. However, deep learning-based wind power forecasting studies have not investigated two aspects. In the first stage, different learning structures considering multiple inputs and multiple outputs have not been discussed. In the second stage, the model extrapolation issue has not been investigated. Therefore, we develop four deep neural networks for the first stage to learn data features considering the input-and-output structure. We then explore the model extrapolation issue in the second stage using different modeling methods. Considering the overfitting issue, we propose a new moving window-based algorithm using a validation set in the first stage to update the training data in both stages with two different moving window processes. Experiments were conducted at three wind farms, and the results demonstrate that the model with single-input–multiple-output structure obtains better forecasting accuracy compared to existing models. In addition, the ridge regression method results in a better ensemble model that can further improve forecasting accuracy compared to existing machine learning methods. Finally, the proposed two-stage forecasting algorithm can generate more accurate and stable results than existing algorithms.

**Keywords**: wind power forecasting, deep neural networks, ensemble learning, extrapolation

# 1 Introduction

## 1.1 Overview

Renewable energy has become a primary focus in academic research and has driven changes in the power industry. By the end of 2020, the world's annual renewable energy generation reached 2799 GW [1]. Among various renewable energy sources, wind energy is a very promising source [2] that takes up the largest proportion [3]. However, the intermittent and dynamic nature of wind energy introduces risks to the economical and reliable operation of power systems [4].

As reported by Muesgens and Neuhoff [5], wind forecasts with 1–4 h ahead of physical energy dispatch are an effective method to reduce the balancing costs caused by wind uncertainties. Thus, numerous studies [6-8] have focused on developing forecasting models using different techniques to improve short-term wind power forecasting (WPF) accuracy. Generally, recent forecasting models can be classified as single model and ensemble model approaches. With single-model approaches, given training data, a statistical modeling or machine learning method is employed to construct a single model to forecast wind power generation [9-11]. In ensemble-model approaches, the forecasting model is developed using an ensemble method to integrate the initial forecasting results from several individual models [13-15]. Most existing ensemble-based studies have used a two-stage framework. For example, Feng et al. [13] exploited different machine learning (ML) techniques to build multiple single forecasting models in the first stage, and then they integrated the results with an algorithm to improve forecasting accuracy. Hao and Tian [15] proposed a two-stage WPF module with a nonlinear ensemble method in the second stage to integrate all components and forecast error values. Ensemble models are more comprehensive; thus, Freedman et al. [12] stated that ensemble models may produce more accurate forecasting results than a single model. However, some factors have not been considered in ensemble-based literature. Thus, in this study, we primarily develop a WPF model in consideration of the following aspects using the two-stage framework.

The first factor we consider is the model input in the first stage. Using hybrid data (e.g., historical wind data and numerical weather prediction (NWP) data) as input has been studied extensively in the literature [10-11] because such data can provide more input information. Nonetheless, historical wind data differ from NWP data, which are generated from weather research forecasting simulation models, in terms of the variance and mean (Section 2). Typically, previous studies have combined historical wind data and NWP data as input to the feature-learning process, which is referred to as single-input learning in this paper. Compared to single-input learning, multiple-input learning uses different channels or methods to learn features to avoid inference between different data types. Therefore, our first goal is to determine whether single-input learning or multiple-input learning is more suitable for the WPF model.

The second factor considered in this paper is the model output structure in the first stage. Typically, previous studies [9-15] have considered the single-output learning structure; however, for deep neural networks, such a learning structure may cause that the gradient cannot be well passed down to the lower layer sometimes. Multiple-output learning has been studied to address this issue [16], and it has been demonstrated that this manner [17] can improve the accuracy of low-level features such that the overfitting issue might be avoided. However, to the best of our knowledge, no previous study that used wind speed and wind power as input [22-23] have considered the multiple-output learning structure when constructing forecasting models. Note that if the input data is of a single type, e.g., historical wind power, the multiple-output learning structure is not suitable. Therefore, considering the application of hybrid data as input, our second goal is to determine whether the multiple-output learning structure can be used to improve forecasting accuracy.

The third factor is the modeling technique employed in the second stage, which blends the forecasting results from the first stage. Dorado-Moreno et al. [21] stated that wind power ramp event forecasting should not depend on long-term historical data. Therefore, we prefer to use the latest data just before the forecasting day as the input to the second-stage ensemble model; however, the data size should not be large. With this method, it is possible to improve forecasting accuracy; however, this may result in model extrapolation considering only a small amount of training data in the second stage. In statistics, model extrapolation is a well-known issue that may occur when examining a model with data that are over the original range [20]. In ensemble-based research, although ML methods, e.g., artificial neural networks (ANN) [15] and support vector regression (SVR) [18], have achieved some success in integrating the results from the first stage, these studies did not consider the model extrapolation issue. Different ML methods have different sensitivity to model extrapolation [28]. When model extrapolation occurs, in some cases, the model shape is kept the same as that at the data boundary. In other case, the original model shape can be ruined at the point extrapolation occurs. Thus, our third goal is to investigate which modeling method is more suitable for the second stage.

## 1.2 Contributions

The two-stage forecasting framework is an effective way to improve WPF accuracy. However, several factors have not been addressed within this framework. To bridge the knowledge gap discussed above, for the first stage, we developed four single forecasting models with different learning structures: single-input–single-output (SISO), multiple-input–single-output (MISO), single-input–multiple-output (SIMO), and multiple-input–multiple-output (MIMO) models. The data used in this study include the historical wind speed, wind power, and NWP data. Even though our goal was to use these four models to effectively learn the data feature from different aspects to provide useful data for the second-stage ensemble model, we also compared the performance of these single forecasting models given a testing dataset. For the second stage, we integrated the results from the first stage using the ridge regression (RR) method to reduce model extrapolation errors [19]. We also exploited three popular machine learning (ML) techniques, i.e., the ANN, SVR, and Gaussian process regression (GPR) techniques, as benchmarks to demonstrate the performance when using RR method in the second stage. In addition, considering the uncertainties and intermittencies of natural wind, we incorporated the two-stage forecasting framework into a moving window based training data updating algorithm. Differing from the moving window algorithm used in the literature [25-26], the proposed algorithm used a validation set in the first stage to adjust the model's parameters. In addition, this algorithm involved two different moving window processes to dynamically update the first-stage and second-stage training data. Finally, we compared the proposed algorithm to several existing algorithms to demonstrate its advantages. Our primary contributions are summarized as follows.

1. We have developed a moving window-based two-stage short-term WPF algorithm.
2. We have developed four deep neural networks in consideration of multiple-input and multiple-output learning structures.
3. We have exploited the RR method to construct an ensemble model at the first time in consideration of the model extrapolation issue in the forecasting process.
4. We have demonstrated the proposed algorithm had a better performance than existing algorithms in numerical experiments.

**1.3 Organization**

The remainder of this paper is organized as follows. Section 2 introduces the first-stage models, and Section 3 describes the moving window-based two-stage forecasting algorithm. Section 4 displays the results of the numerical studies. Finally, Section 5 provides the paper conclusion.

**2 Development of first-stage models**

In the deep learning-based WPF model, the combination of convolutional neural networks (CNN) and long short-term memory (LSTM) networks has shown outstanding performance [9-11]. Therefore, we exploit such a combination to construct the SISO, SIMO, MISO, and MIMO models in the first stage. Table 1 describes the variables used in this paper.

Table 1. Nomenclature.

| | | | | | |
|---|---|---|---|---|---|
| $h$ | Prediction horizon | $L_{p,s}$ | Loss function of the combination of wind power and speed | $\hat{Y}$ | Final power forecasting results |
| $x_p$ | Vector of historical wind power data | $\beta$ | Weight of wind speed in $L_{p,s}$ | $X_{t2}$ | input data of $S_{t2}$. |
| $l_{c1}$ | Stage 1 update cycle, which is defined as an integral multiple of $l_{c2}$ | $M_\gamma$ | SIMO model with output $\hat{Y}_\gamma$ | $X_{te}$ | Input data of $S_{te}$ |
| $Y_{t1}$ | Real power observation of $S_{t1}$. | $M_\delta$ | SISO model with output $\hat{Y}_\delta$ | $\hat{Y}_\beta$ | Power forecasting results of MISO model |
| $\varphi_{p_t}$ | Real wind power at time $t$ | $N$ | Number of testing points | $\hat{Y}_\delta$ | Power forecasting results of SISO model |
| $x_w$ | Vector of NWP data | $t_1$ | Starting point of $S_{t1}$ | $\alpha$ | Weight of wind power and in loss function $L_{p,s}$ |
| $\varphi_{p_t}$ | Historical wind power at time $t$ | $t_2$ | Starting point of $S_{t2}$ | $M_\alpha$ | MIMO model with output $\hat{Y}_\alpha$ |
| $\hat{\varphi}_{s_t}$ | Predicted wind speed at time $t$ | $Y_{v1}$ | Real power observation of $S_{v1}$. | $M_\beta$ | MISO model with output $\hat{Y}_\beta$ |
| $S_{t1}$ | First stage training set | $M_\beta$ | MISO model with output $\hat{Y}_\beta$ | $M_{RR}$ | RR model with output $\hat{Y}$ |
| $S_{t2}$ | Second stage training set | $M_\gamma$ | SIMO model with output $\hat{Y}_\gamma$. | $Y$ | Real wind power generation |
| $L_p$ | Loss function of wind power | $\varphi_{w_t}$ | NWP data at time $t$ | $t_v$ | Starting point of $S_{v1}$ |
| $X_{t1}$ | input data of $S_{t1}$ | $\hat{\varphi}_{p_t}$ | Power prediction at time $t$ | $t_e$ | Starting point of $S_{te}$ |
| $X_{v1}$ | input data of $S_{v1}$. | $m$ | Number of data used for modeling | $M_\alpha$ | MIMO model with output $\hat{Y}_\alpha$ |
| $\hat{Y}_\alpha$ | Power forecasting results of MIMO model | $S_{v1}$ | First stage validation set | $M_\delta$ | SISO model with output $\hat{Y}_\delta$ |
| $\hat{Y}_\gamma$ | Power forecasting results of SIMO model | $S_{te}$ | Testing set | $M_{RR}$ | RR model with output $\hat{Y}$ |

## 2.1 SISO model

The SISO model (Fig. 1) follows a structure that has been used extensively in the literature, where the input data are combined as input to the forecasting model, and single forecasting results are output.

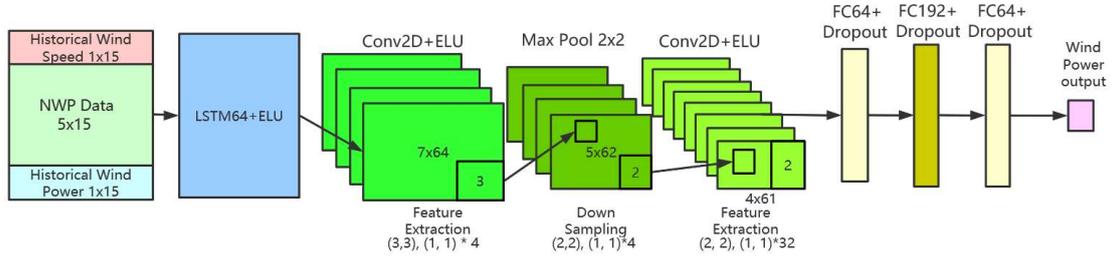

Fig. 1. SISO model architecture using deep neural networks.

Due to the prediction horizon $h$ and time steps $n$, if we are at time $t$, we use the historical wind power data $x_p = \{\varphi_{p_{t-n}}, \ldots, \varphi_{p_{t-1}}, \varphi_{p_t}\}$, historical wind speed data $x_s = \{\varphi_{s_{t-n}}, \ldots, \varphi_{s_{t-1}}, \varphi_{s_t}\}$, and NWP data $x_w = \{\varphi_{w_{t-h+1}}, \ldots, \varphi_{w_{t+h}}, \varphi_{w_{t+h+1}}, \ldots, \varphi_{w_{t+2h-1}}\}$ as the input to construct the forecasting model. Here, the NWP data are obtained one day in advance such that the settings using the NWP data after the forecasting time point are reasonable. The SISO model strategy is employed to solve the regression problem by minimizing the loss function in each batch of outputs, as given in Eq. (1).

$$L_p = \frac{1}{m}\left(\sum \left\|\hat{\varphi}_{p_{t+h}} - \varphi_{p_{t+h}}\right\|^2\right) \qquad (1)$$

Fig. 1 shows the historical wind speed, historical wind power, and NWP data combined into a 7 × 15 matrix, where digit 7 represents the seven input variables, i.e., the five NWP variables (wind speed, wind direction, humidity, air pressure, and temperature), the historical wind speed variable, and the historical wind power variable. For example, if we are at 8:00 AM, for the NWP data, 15 represents the data from 8:15 AM to 11:45 AM. For the wind speed and power data, 15 represents the data from 4:30 AM to 8:00 AM. The constructed matrix is accepted by a 64 hidden-size LSTM layer. The outputs are then passed to the CNN layers with the ELU activation function in the first layer. Taking the first convolutional layer as an example, 7 × 64 represents the input size, there are four kernels, (3, 3) is the kernel size, and (1, 1) is the step length. The wind power is output at the last fully-connected (FC) layer after three FC layers.

## 2.2 SIMO model

Figure 2 shows the SIMO model architecture. Here, we assume that the multiple-output learning structure can potentially avoid the overfitting issue to construct a more robust model. Compared to the SISO model, the SIMO model forecasts wind speed and wind power (Fig. 2). Thus, the loss function can be expressed as follows.

$$L_{p,s} = \frac{1}{m}\left(\alpha\Sigma\left|\left|\hat{\varphi}_{p_{t+h}} - \varphi_{p_{t+h}}\right|\right|^2\right) + \frac{1}{m}\left(\beta\Sigma\left|\left|\hat{\varphi}_{s_{t+h}} - \varphi_{s_{t+h}}\right|\right|^2\right) \qquad (2)$$

The Pearson correlation coefficient between the real wind speed and wind power is greater than 0.933 in the three selected wind farms; thus, we set $\alpha$ to 1 and $\beta$ to 0.9.

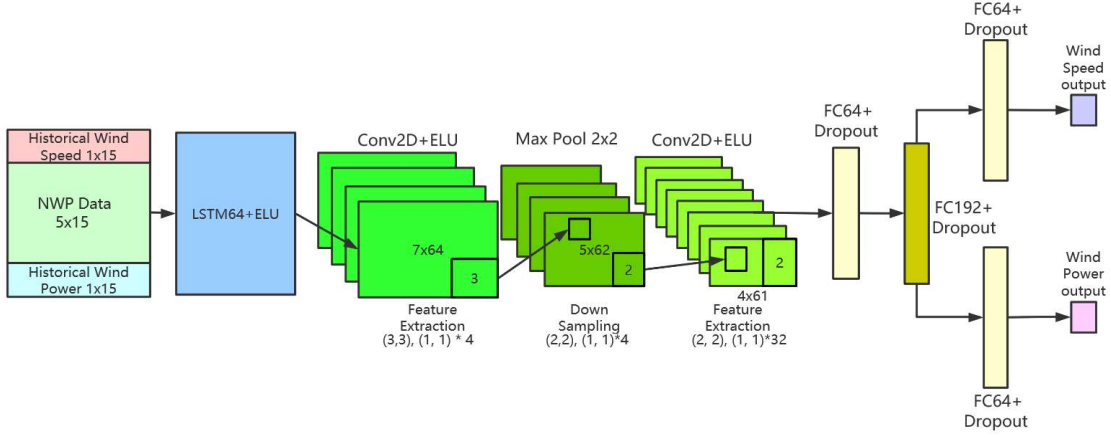

Fig. 2. SIMO model architecture using deep neural networks.

**2.3 MISO model**

A comparison of the NWP wind speed, real wind speed, and real wind power data is shown in Fig. 3. From such marginal plots, the differences between these three data appear significant; however, based on the trend of each datum type, the real wind power data have a more similar trend with the real wind speed data than the NWP wind speed data. In addition, the real wind speed exhibits some similar patterns to the NWP wind speed. Thus, for the MISO model, we assume that the multiple-input structure can potentially learn features unaffected by other types of data to improve forecasting accuracy. Figure 4 shows the MISO model architecture. Here, the SISO structure is used extract the features of the NWP data; however, for historical wind speed and wind power data, we employ another LSTM layer to learn their internal patterns. The features extracted from the three different types of input data are then concatenated before being output.

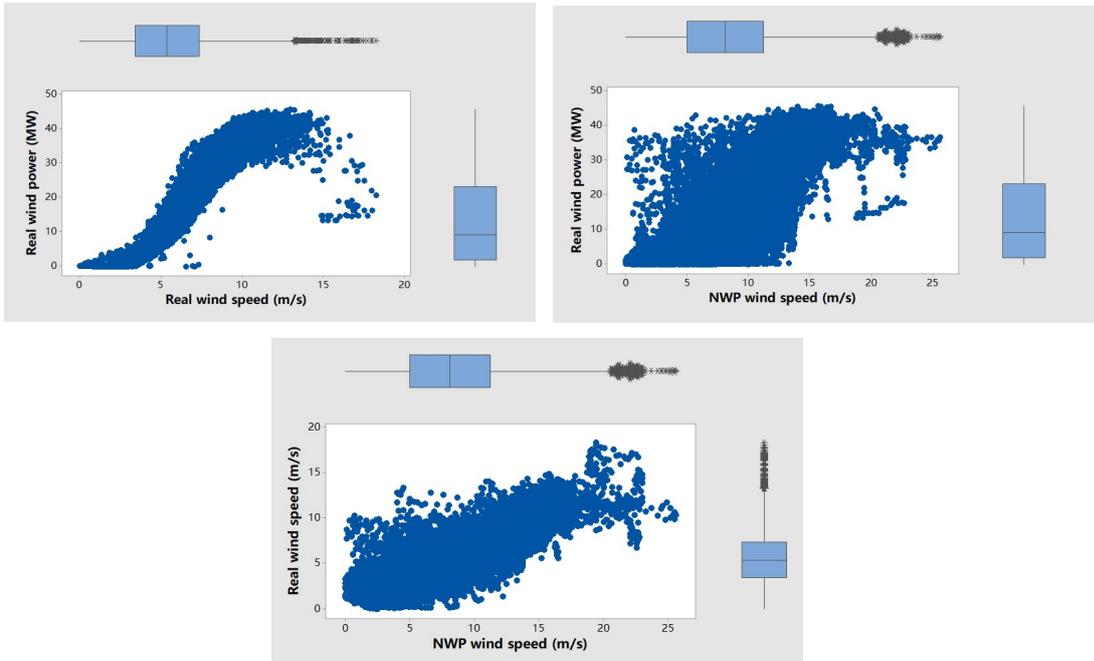

Fig. 3. Marginal plot between real wind power, real wind speed, and NWP wind speed.

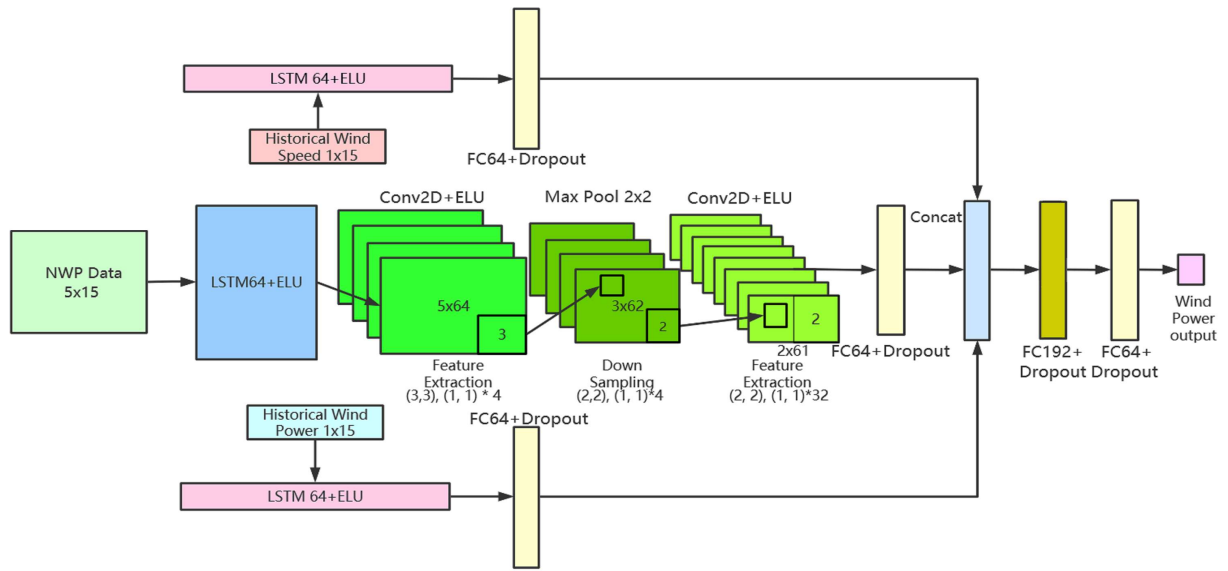

Fig. 4. Architecture of MISO model using deep neural networks.

## 2.4 MIMO model

The structure of the MIMO model integrates the MISO and SIMO models (Fig. 5). If the assumptions made for the MISO and SIMO models are true, such a comprehensive MIMO model is assumed to further improve the forecasting accuracy. However, if only one of the aforementioned two assumptions is true, predicting the MIMO model performance will be difficult. Although the MIMO model structure is more complicated compared to the other three models, there is no guarantee that the MIMO model will produce the best forecasting results.

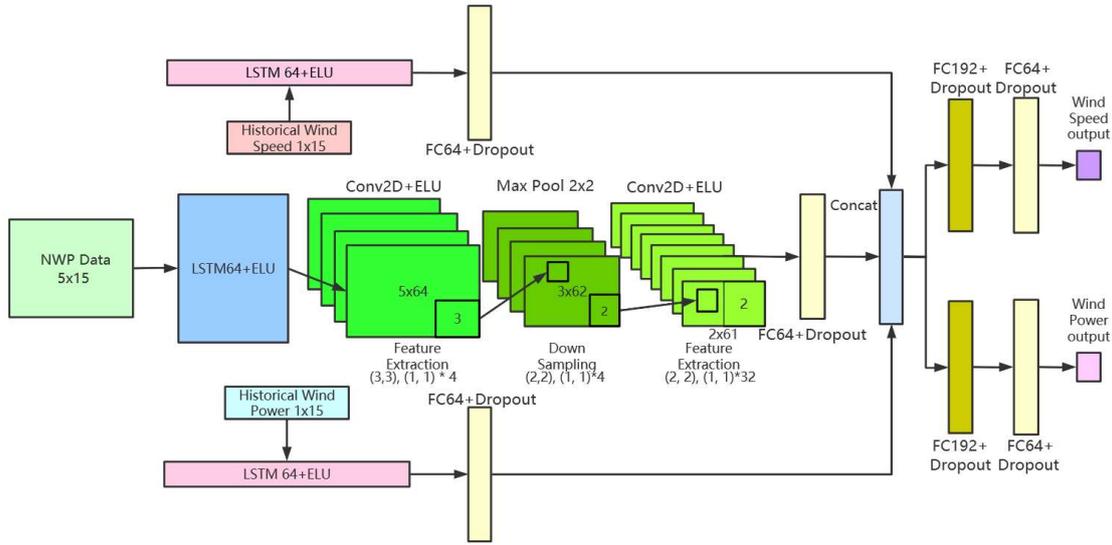

Fig. 5. Architecture of the MIMO model using deep neural networks.

As shown in Figs. 1-2 and Figs. 4-5, we attempt to provide a fair comparison using the same combination of CNN and LSTM layers and the same parameters in each layer. We determined the number of CNN and LSTM layers and the parameters in each layer based on our preliminary experiment, and this combination obtained a better forecasting accuracy than using SVR and GPR. However, we did not focus on optimizing the best combination of CNN and LSTM layers because this was beyond the scope of the study. Thus, with these four structures, we can learn the data feature from different perspectives and simultaneously investigate model performance by changing only the input-and-output structure.

## 3 Two-stage forecasting algorithm

The data used for each stage are critical relative to developing a two-stage forecasting algorithm. Considering the overfitting issue, we ensured that the training sets in the first and second stages did not overlap. In the first stage, we considered a validation set to adjust the parameters of the training model; however, we did not use the cross-validation method to obtain a generally good model. The wind data, especially the NWP data in consecutive days, demonstrate some similarities [24]; thus, such a design would result in that the model performing well on the validation set possibly has a good performance on the testing set. Therefore, we used the data before the forecasting day as the validation set. In the second stage, the ensemble model was employed to further improve the forecasting accuracy; thus, we used latest data before the forecasting day in the training set. Compared to the data used in the first stage, the second stage only required a small amount of data. Note that the second-stage training data may not necessarily be the same as the validation set because sometimes the wind changed unpredictably, and such a difference would help avoid overfitting. Even though such settings in the second stage would benefit forecasting accuracy; the sudden change of testing wind data would be beyond the range of the second-stage training data. As a result, the model extrapolation may occur. Here, if the second-stage model construction method is not well selected, the forecasting accuracy would be seriously affected by the model extrapolation issue.

Fig. 6 illustrates the proposed moving window-based two-stage forecasting algorithm. The proposed algorithm involves two moving window processes to the training datasets update. Here, in a certain time period, Moving Window 1 moves the training and testing datasets forward. We divided the testing datasets into daily subsets; thus, Moving Window 2 was used to move the second-stage training set forward as the testing subset evolved. Moving Window 2 was embedded into Moving Window 1. For this design, Moving Window 1 provides a generally good forecasting model, and Moving Window 2 considers more recent information to reduce forecasting errors in the first-stage models. We then set the window size of the first-stage training set to one year. In addition, the window sizes of the validation, second-stage training, and testing sets were set to 10 days. For example, when forecasting the short-term wind power on January 11, 2020, we used data from 2019 as the first-stage data and data from January 1, 2020 to January 10, 2020 as the validation and second-stage training sets. When forecasting the wind power on January 20, 2020, we used the 2019 data as the first-stage data and data from January 1, 2020 to January 10, 2020 as the validation set. However, the second-stage training data were updated to include date from January 10, 2020 to January 19, 2020.

In addition to the moving window algorithms, the four proposed deep neural networks (Section 2) were applied to the first-stage data to adjust parameters. In the second stage, the first-stage models (i.e., MIMO, MISO, SIMO, and SISO) were applied directly to the second-stage wind data to obtain the forecasting results of $\hat{Y}_\alpha, \hat{Y}_\beta, \hat{Y}_\gamma$, and $\hat{Y}_\delta$ (see Table 1 for the corresponding definitions). This was a key step in the two-stage forecasting framework because it connected the first and second stages. As a result, it was possible to learn the forecasting errors from the first stage models by comparing $\hat{Y}_\alpha, \hat{Y}_\beta, \hat{Y}_\gamma$, and $\hat{Y}_\delta$ to their corresponding real power $Y_{t2}$. Then, an ensemble model generated by RR method was constructed with $\hat{Y}_\alpha, \hat{Y}_\beta, \hat{Y}_\gamma$, and $\hat{Y}_\delta$ as input and $Y_{t2}$ as output.

When forecasting the next-period wind power, the testing data were first input to the four well-trained models in the first stage to obtain $\hat{Y}'_\alpha, \hat{Y}'_\beta, \hat{Y}'_\gamma$, and $\hat{Y}'_\delta$, which were then input to the RR model (see the blue lines in Fig. 6). Algorithm 1 shows the detailed information about this two-stage forecasting algorithm.

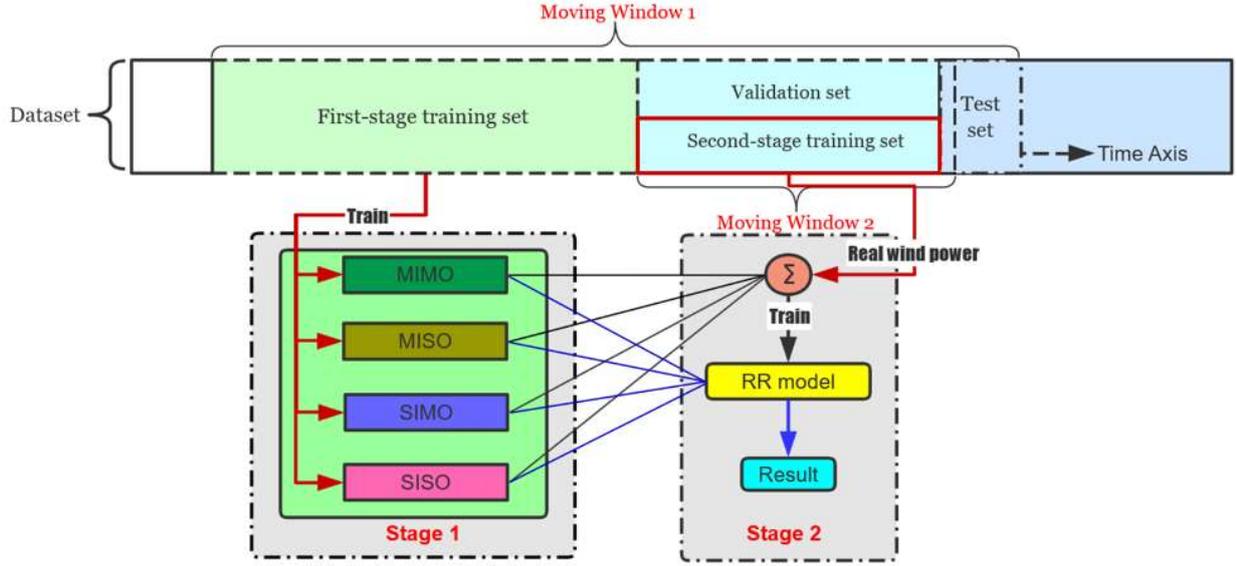

Fig. 6. Proposed two-stage forecasting algorithm.

| Algorithm 1. Proposed two-stage forecasting algorithm |
| --- |
| Require: $S_{t1}, S_{t2}, S_{v1}, S_{te}, t_1, t_2, t_v, t_e, X_{t1}, X_{t2}, X_{v1}, X_{te}, l_{c1}, l_{c2}, M_\alpha, M_\beta, M_\gamma, M_\delta, M_{RR}, Y_{t1}, Y_{t2}, Y_{v1}$, |
| 1: while $S_{te}$ is not null: |
|     2: do train $M_\alpha, M_\beta, M_\gamma, M_\delta$ with input $X_{t1}$ and output $Y_{t1}$ as training set, and validate them with input $X_{t1}$ and output $Y_{v1}$. |
|     3: for $i \leftarrow 1$ to $(\frac{l_{c1}}{l_{c2}})$: |
|         4: do collect $\hat{Y}_\alpha \leftarrow M_\alpha(X_{t2}), \hat{Y}_\beta \leftarrow M_\beta(X_{t2}), \hat{Y}_\gamma \leftarrow M_\gamma(X_{t2}), \hat{Y}_\delta \leftarrow M_\delta(X_{t2})$ |
|         5: train $M_{LR}$ with $\hat{Y}_\alpha, \hat{Y}_\beta, \hat{Y}_\gamma, \hat{Y}_\delta$ as input and $Y_{t2}$ as output |
|         6: for $j \leftarrow 1$ to $l_{c2}$: |
|             7: do collect $\hat{Y}'_\alpha[j] \leftarrow M_\alpha(X_{te}[j]), \hat{Y}'_\beta[j] \leftarrow M_\beta(X_{te}[j]), \hat{Y}'_\gamma[j] \leftarrow M_\gamma(X_{te}[j]), \hat{Y}'_\delta[j] \leftarrow M_\delta(X_{te}[j])$ |
|             8: get RESULT: $\hat{Y}'_{te}[j] \leftarrow M_{LR}(\hat{Y}'_\alpha[j], \hat{Y}'_\beta[j], \hat{Y}'_\gamma[j], \hat{Y}'_\delta[j])$ |
|         9: $t_v \leftarrow t_v + l_{c2}, t_e \leftarrow t_e + l_{c2}$ |
|         10: update $S_{t2}, S_{te}$ |
|     11: $t_1 \leftarrow t_1 + l_{c1}$ |
|     12: $t_2 \leftarrow t_2 + l_{c1}$ |
|     13: update $S_{t1}, S_{t2}$ |
| 14: end while |

## 4 Case studies

We implemented the proposed algorithm with the data from three wind farms in China to evaluate the short-term WPF accuracy. Table 2 shows the basic information of these wind farms. The wind conditions of the three wind farms were not quite the same when comparing the mean and standard deviations of their wind speed and power. Therefore, the results of our case studies are likely to demonstrate the generalizability of the forecasting models.

The wind data for 2018 and 2019 were used in these case studies. Here, we randomly selected the 10-day data from each season as a testing set to examine the performance of the models. Thus, when Moving Window 1 moved to the new step, the evaluation set for the first day of forecasting coincided with the second-stage training set.

Table 2. Basic wind farm information.

| No. of wind farm(s) (WF) | WF 1 | WF 2 | WF 3 |
|---|---|---|---|
| Number of wind turbines | 33 | 24 | 24 |
| Installed capacity | 49.5 MW | 48 MW | 48 MW |
| Type of manufacturer | Envision EN70-1500 | Vestas V80-2000 | Gamesa G90 |
| Hub-height of the tower(s) | 70 m | 80 m | 90 m |
| Cut-in/cutout wind speed | 4.0 m/s / 25 m/s | 3.5 m/s / 25 m/s | 3 m/s / 21 m/s |
| Rated wind speed | 11.6 m/s | 14.5 m/s | 11 m/s |
| Swept area | 3915 m$^2$ | 5027 m$^2$ | 6362 m$^2$ |
| Mean wind speed | 5.38 m/s | 5.73 m/s | 6.25 m/s |
| Mean wind power | 12.73 MW | 12.91 MW | 12.07 MW |
| Standard deviation of wind speed | 2.85 | 3.33 | 3.53 |
| Standard deviation of wind power | 12.89 | 12.87 | 13.51 |

The prediction for the 2-h ahead wind power was typically used as a benchmark to evaluate the accuracy of a forecasting model in the current power system. Thus, we implemented the proposed algorithm to forecast the 2-h ahead wind power at the target wind farms and used the root mean square error (RMSE) and mean absolute error (MAE) to evaluate the performance of the forecasting results.

$$RMSE = \sqrt{\frac{1}{N}(\sum_{i=1}^{N}(Y_i - \hat{Y}_i)^2} \quad (3)$$

$$MAE = \frac{1}{N}\sum_{i=1}^{N}|Y_i - \hat{Y}_i| \quad (4)$$

**4.1 Experiment 1**

In the first experiment, we investigated the effects of the four models on forecasting accuracy in consideration of the different model structures in the first stage. Here, Moving Window 2 (Fig. 6) was not performed in this experiment. In other words, we only constructed the first-stage models and used them to forecast the 10-day wind power. Then, we obtained the results for the four seasons (Table 3) after applying the evaluation measurement in Eqs. (3) and (4). The best forecasting accuracy in each row of Table 3 is highlighted in bold.

The results given in Table 3 demonstrate that the accuracies of the proposed models differed in different seasons, which indicates that the different input-and-output learning structures generated different forecasting results. In addition, we found that most results from the SIMO model demonstrated the best forecasting accuracy. The average performance of these four models is shown in Fig. 7.

Table 3. Comparison of forecasting results.

| | Spring | MIMO | MISO | SIMO | SISO | | Summer | MIMO | MISO | SIMO | SISO |
|---|---|---|---|---|---|---|---|---|---|---|---|
| WF1 | RMSE | 5.5242 | 5.5795 | **5.3376** | 5.7981 | WF1 | RMSE | 3.6307 | 3.7707 | **3.5861** | 3.8215 |

| | | MIMO | MISO | SIMO | SISO | | | MIMO | MISO | SIMO | SISO |
|---|---|---|---|---|---|---|---|---|---|---|---|
| | MAE | 4.2093 | 4.2734 | **4.0671** | 4.4994 | | MAE | 2.0724 | 2.2217 | **2.0469** | 2.3701 |
| WF2 | RMSE | 2.9125 | 2.925 | **2.5205** | 2.704 | WF2 | RMSE | 4.3182 | 4.3391 | **4.0855** | 4.5418 |
| | MAE | 1.8784 | 1.8315 | **1.6091** | 1.7583 | | MAE | 3.2826 | 3.3305 | **3.1694** | 3.5356 |
| WF3 | RMSE | 2.3081 | 2.2459 | **2.0235** | 2.2106 | WF3 | RMSE | 5.4227 | 5.567 | **4.7035** | 5.1366 |
| | MAE | 1.3194 | 1.4657 | **1.1654** | 1.4309 | | MAE | 4.0186 | 4.1316 | **3.5202** | 3.8857 |
| Autumn | | MIMO | MISO | SIMO | SISO | Winter | | MIMO | MISO | SIMO | SISO |
| WF1 | RMSE | 5.2346 | 5.2490 | **4.9586** | 5.0682 | WF1 | RMSE | 5.0679 | 5.2256 | **4.8618** | 5.1487 |
| | MAE | 3.8446 | 3.8977 | **3.6221** | 3.8527 | | MAE | 3.5209 | 3.73 | **3.3299** | 3.6106 |
| WF2 | RMSE | 4.0866 | 4.1866 | **3.8224** | 4.0557 | WF2 | RMSE | 4.7144 | **4.7109** | 4.7249 | 4.8591 |
| | MAE | 2.9126 | 3.0005 | **2.7651** | 2.9071 | | MAE | 3.3419 | 3.3122 | **3.3112** | 3.5154 |
| WF3 | RMSE | 5.2954 | 5.2845 | **4.5240** | 4.6975 | WF3 | RMSE | **4.9402** | 5.0171 | 4.9409 | 5.1371 |
| | MAE | 3.5466 | 3.4424 | **2.9324** | 3.1839 | | MAE | **3.7694** | 3.8224 | 3.8051 | 3.9060 |

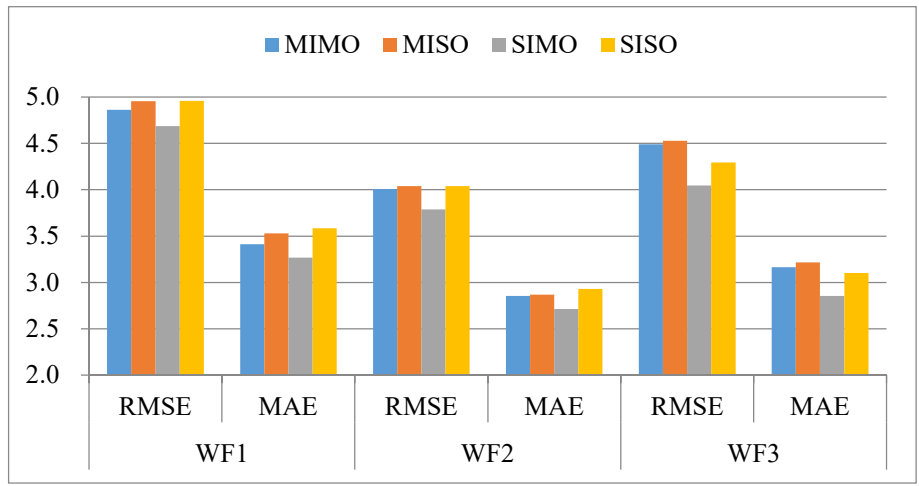

Fig. 7. Average forecasting accuracy of proposed forecasting models.

Fig. 7 shows that SIMO model had a better forecasting performance than the other three models. The MIMO model demonstrated a slightly better performance than the MISO model in these three cases, indicating that the multiple-output structure was effective in terms of improving the forecasting accuracy. However, when comparing MISO to SISO, we cannot obtain a clear conclusion that the multiple-input structure would improve forecasting accuracy. Thus, these results indicated that mixing the different input data together as a single input was better when learning the data features. In addition, adding an auxiliary output was able to help improve forecasting accuracy. Fig. 8 shows the variances of the absolute difference between the predicted and real wind powers obtained by four proposed models. We found that the differences were not quite the same in each case. However, the variances of the SIMO model were the lowest in all cases, indicating that the SIMO model generated more consistently stable results than the other models. To further examine these conclusions obtained from Figs. 7 and 8, we combined all forecasting accuracies for the four seasons in Table 3 and performed a statistical analysis using the paired *t*-test. As shown in Table 4, all *p*-values were less than 0.05, and all the confidence intervals (CI) were less than zero when comparing the forecasting accuracies of the SIMO model to those of the other models. In other words, the SIMO model generated statistically better forecasting accuracies than the other models. As shown in Table 4, even though the SISO model statistically had the same performance as the MIMO and MISO models, we used all four models in the first stage to produce the forecasting results for the second stage because the statistically same forecasting accuracy did not indicate that they output the same WPF results. This comparison was performed to emphasize the outstanding performance of SIMO model.

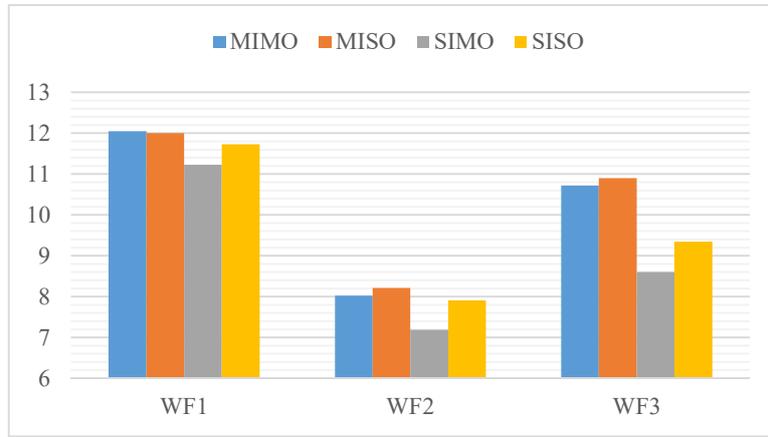

Fig. 8. Variances of absolute difference between predicted and real wind powers from the four proposed models.

Table 4. Statistical analysis of forecasting accuracies using paired *t*-test ($\alpha=0.05$).

|  | SIMO vs. SISO | SIMO vs. MISO | SIMO vs. MIMO | SISO vs. MISO | SISO vs. MIMO | MISO vs. MIMO |
| --- | --- | --- | --- | --- | --- | --- |
| $p_{rmse}$ | <0.000 | 0.001 | 0.002 | 0.312 | 0.764 | 0.023 |
| $CI_{rmse}$ | (-0.3369, -0.1781) | (-0.4955, -0.1731) | (-0.4387, -0.1223) | (-0.2363, 0.0826) | (-0.1881, 0.142) | (0.0091, 0.0984) |
| $p_{MAE}$ | <0.000 | <0.000 | 0.004 | 0.995 | 0.301 | 0.035 |
| $CI_{MAE}$ | (-0.3235, -0.1952) | (-0.3738, -0.1454) | (-0.3183, -0.0771) | (-0.1091, 0.1084) | (-0.0632, 0.1864) | (0.0051, 0.1187) |

($p$ indicates the $p$-value; $CI$ represents the confidence interval)

### 4.2 Experiment 2

We implemented the proposed two-stage algorithm to construct the forecasting model. Although the SIMO model discussed in Section 4.1.1 demonstrated better performance compared to the other models, the second stage of the algorithm attempts to further improve accuracy using the RR method. As mentioned previously, we preferred to use the latest information before the forecasting day to construct the ensemble model. This method effectively updates the ensemble model over time; however, this may lead to model extrapolation caused by the quantity of the input data. Thus, the selecting the most appropriate modeling technique in the second stage is critical. Here, we used the SVR
13

with a radial basis function (RBF) kernel, ANN, and GPR methods as benchmarks to better demonstrate the RR method performance. In addition, we used the *sklearn* toolbox in Python 3.7.3 to implement these four methods and tuned the parameters of each method using the grid search algorithm. The parameter settings are listed in Table 5.

Table 5. Parameter settings and tuning of the four modeling methods.

| Model | Parameters | Search range |
| --- | --- | --- |
| RR | Fit intercept = False | Alpha = [1:100], stepsize = 1 |
| SVR | Kernel = RBF | C = [0.5:20], stepsize = 0.5 |
| ANN | Hidden layer size = (64,128), activation = ReLU, solver = Adam | Alpha = [0.0001, 0.0003, 0.001, 0.003, 0.01] |
| GPR | Kernel = RBF | Alpha = [0.1:1], stepsize = 0.1 |

Table 6 shows the forecasting accuracies for each season using different statistical modeling techniques, and Fig. 9 shows the interval plot of the forecasting accuracies presented in Table 6. In Table 6 and Fig. 9, we also showed the forecasting accuracies obtained by the SIMO model to demonstrate the advantages of the ensemble model. Although the ensemble models with the four statistical modeling methods did not outperform the SIMO model in all cases, the RR model outperformed the SIMO model in each case. In addition, the 95% CI for RR method was better than all of them, which indicated that the RR method was statistically likely to generate better forecasting results. Compared to the RR method, the ML techniques did not exhibit better performances in the second-stage model. In certain cases, the SVR- and GPR-based ensemble models demonstrated considerably poorer performances than the other models (marked as red in Table 6). We further investigated this observation by plotting the forecasting results of the WF3 winter case, which are shown in Fig. 10.

Table 6. Forecasting accuracy using different modeling techniques for ensemble.

| | Spring | SIMO | RR | SVR | ANN | GPR | | Summer | SIMO | RR | SVR | ANN | GPR |
| --- | --- | --- | --- | --- | --- | --- | --- | --- | --- | --- | --- | --- | --- |
| WF1 | RMSE | 5.3376 | **5.2612** | 5.4061 | 5.2697 | 5.4689 | WF1 | RMSE | 3.5861 | 3.4939 | **3.4281** | 3.5430 | 3.4691 |
| | MAE | 4.0671 | **4.0077** | 4.0857 | 3.9931 | 4.2199 | | MAE | 2.0469 | 1.9085 | **1.8168** | 2.0917 | 1.9829 |
| WF2 | RMSE | 2.5205 | **2.4865** | 2.4887 | 2.5469 | 2.5748 | WF2 | RMSE | 4.0855 | **3.9674** | 4.0611 | 3.9793 | 4.1710 |
| | MAE | 1.6091 | 1.5803 | **1.4260** | 1.5800 | 1.5519 | | MAE | 3.1694 | **3.0383** | 3.1411 | 3.0375 | 3.1980 |
| WF3 | RMSE | 2.0235 | **2.0183** | 2.0561 | 2.1224 | 2.1118 | WF3 | RMSE | 4.7035 | 4.6511 | **4.5685** | 4.7544 | 4.6760 |
| | MAE | 1.1654 | 1.1502 | **1.0217** | 1.1678 | 1.1285 | | MAE | 3.5202 | 3.4728 | **3.3738** | 3.5205 | 3.4811 |
| | Autumn | SIMO | RR | SVR | ANN | GPR | | Winter | SIMO | RR | SVR | ANN | GPR |
| WF1 | RMSE | 4.9586 | **4.7464** | 4.8681 | 4.8041 | 4.8259 | WF1 | RMSE | 4.8618 | 4.7531 | 4.8418 | **4.6949** | 5.2367 |
| | MAE | 3.6221 | **3.5090** | 3.5100 | 3.6443 | 3.5612 | | MAE | 3.3299 | 3.2343 | 3.2055 | **3.1636** | 3.6595 |
| WF2 | RMSE | 3.8224 | **3.6748** | 4.2732 | 3.8029 | 5.1703 | WF2 | RMSE | 4.7249 | **4.5551** | 4.8481 | 4.6041 | 5.1104 |
| | MAE | 2.7651 | **2.6277** | 2.9386 | 2.7373 | 3.3129 | | MAE | 3.3112 | **3.2276** | 3.3947 | 3.2898 | 3.7321 |
| WF3 | RMSE | 4.5240 | **4.3834** | 4.5486 | 4.4422 | 5.3223 | WF3 | RMSE | 4.9409 | **4.6874** | 6.2021 | 4.9197 | 7.6411 |
| | MAE | 2.9324 | **2.8778** | 2.9795 | 2.9548 | 3.3843 | | MAE | 3.8051 | **3.5662** | 4.3800 | 3.7487 | 4.9461 |

(Lowest RMSE and MAE values in each case are shown in bold)

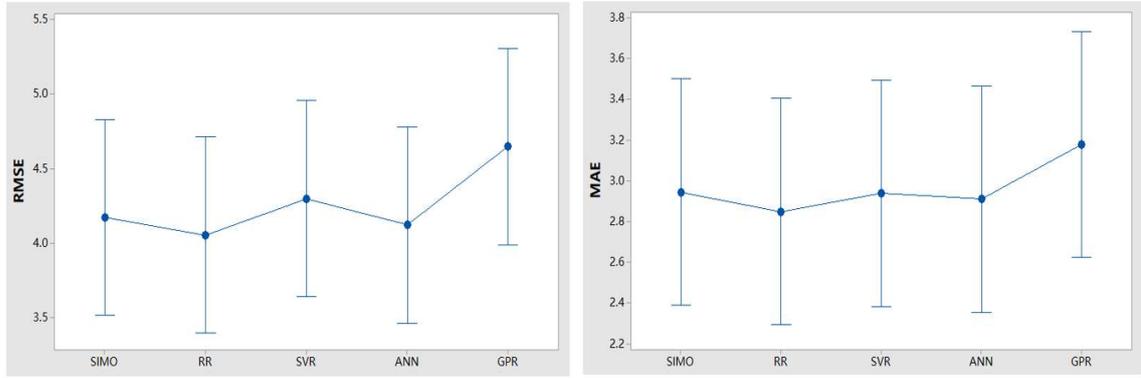

Fig. 9. 95 % confidence interval plot of forecasting accuracies using different modeling techniques.

Fig. 10 shows that all four forecasting methods demonstrated similar performances, except in time steps 729–785. The SVR- and GPR-based ensemble models obtained very different forecasting results (red circle in the figure) compared to the others in time steps 729–785. Such irrational forecasting phenomenon occurred on the eighth forecasting day; thus, we explored the input information used to construct the eighth ensemble model and input information to examine the eighth model, as shown in Fig. 11, where the first 1000 data were used for the ensemble model construction, and the rest were used for testing. The range of the input data when the ensemble model was constructed was between 0 and 38 MW, and the remaining data had some wind power over 38 MW (dashed rectangle, Fig. 11). In this case, the SVR- and GPR-based models may be very sensitive to the extrapolation issue such that the forecasting results could become much worse in testing day eight. For comparison, the RR and ANN methods showed good performance when model extrapolation occurred, which demonstrated both the advantage of the RR as an ensemble modeling technique and indicated the importance of modeling technique selection considering the extrapolation issue.

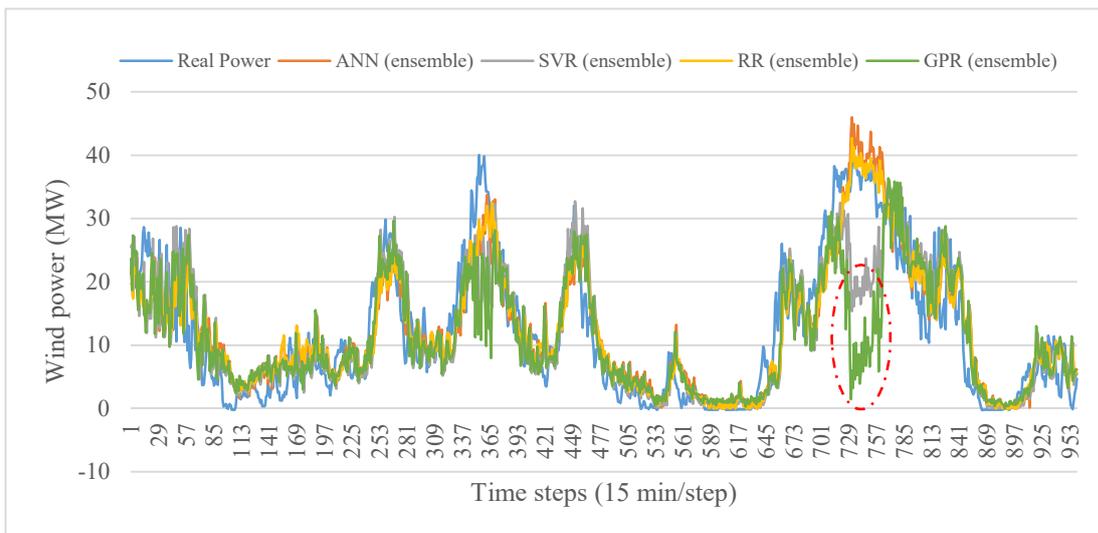

Fig. 10. Ten-day forecasting results in winter case of WF 3 using different ensemble techniques.

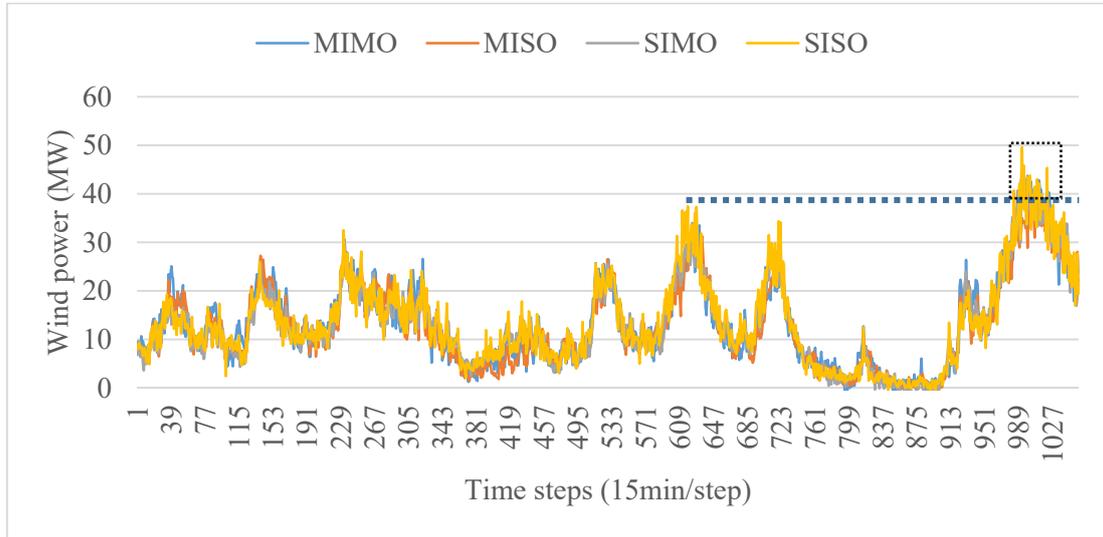

Fig. 11. Input information for eighth day forecasting in the winter case of WF 3 (the first 1000 data were used for modeling, and the rest were used for testing).

**4.3 Experiment 3**

In this experiment, we exploited five existing methods as benchmarks to compare the performance of the proposed forecasting algorithm. Here, we used the persistence (P) and auto regressive integrated moving average (ARIMA) models, which have been used extensively in the literature as benchmarks for short-term wind power/speed forecasting. In addition, we utilized two two-stage forecasting models from Feng et al. [13] and Liu et al. [27]. The MM-DFS method by Feng et al. [13] used multiple single models in the first stage and deep feature selection techniques in the second stage, and the DWT-LSTM method by Liu et al. [27] used discrete wavelet transform and LSTM techniques. In addition, we took advantage of the smart deep learning based wind forecasting method (SDL) by Liu et al. [11], which utilized a combination of CNN and LSTM networks, as another benchmark.

Table 7 shows the forecasting accuracy results. As seen, the proposed two-stage forecasting (TSF) algorithm demonstrated the lowest RMSE and MAE values in all cases. In addition, Fig. 12 shows the CI of the forecasting results obtained using different forecasting algorithms, and the TSF algorithm obtained the best forecasting accuracies. In addition, we integrated all forecasting results and statistically analyzed the absolute difference between the forecasting results and the real wind power. Table 8 shows that the proposed algorithm obtained the lowest mean and variance values, which indicates that the proposed algorithm could generate more accurate and stable results compared to the other methods. An interesting finding is all of these two-stage algorithms (the proposed algorithm, MM-DFS, and DWT-LSTM) obtained overall better forecasting results than the other three because ensemble-based models are more comprehensive.

Table 7. Forecasting accuracy of the proposed two-stage forecasting (TSF) algorithm and existing methods.

| Spring | TSF | P | MM-DFS | DWT-LSTM | ARIMA | SDL | Summer | TSF | P | MM-DFS | DWT-LSTM | ARIMA | SDL |
|---|---|---|---|---|---|---|---|---|---|---|---|---|---|

|     |      | TSF    | P      | MM-DFS | DWT-LSTM | ARIMA  | SDL    |        |      | TSF    | P      | MM-DFS | DWT-LSTM | ARIMA  | SDL    |
| --- | ---- | ------ | ------ | ------ | -------- | ------ | ------ | ------ | ---- | ------ | ------ | ------ | -------- | ------ | ------ |
| WF1 | RMSE | **5.2612** | 5.9184 | 5.6352 | 5.5617 | 6.1944 | 5.8643 | WF1 | RMSE | **3.4939** | 4.0955 | 3.9246 | 3.8153 | 4.4910 | 4.0972 |
|     | MAE  | **4.0377** | 4.3259 | 4.2286 | 4.1690 | 4.6378 | 4.4115 |     | MAE  | **1.9085** | 2.0696 | 2.1012 | 2.1562 | 2.2277 | 2.3686 |
| WF2 | RMSE | **2.4865** | 2.8930 | 2.6822 | 2.6369 | 2.9794 | 2.9825 | WF2 | RMSE | **3.9674** | 4.4904 | 4.3652 | 4.1982 | 4.6609 | 4.3447 |
|     | MAE  | **1.5803** | 1.6245 | 1.6263 | 1.6514 | 1.6854 | 1.8894 |     | MAE  | **3.0383** | 3.3055 | 3.2568 | 3.2341 | 3.4266 | 3.2521 |
| WF3 | RMSE | **2.0483** | 2.3462 | 2.3082 | 2.4391 | 2.4515 | 2.5972 | WF3 | RMSE | **4.6511** | 5.9998 | 4.7033 | 5.0331 | 6.3524 | 5.8524 |
|     | MAE  | **1.1502** | 1.2481 | 1.2576 | 1.3620 | 1.2827 | 1.6682 |     | MAE  | **3.4728** | 4.2760 | 3.5079 | 3.8447 | 4.6812 | 4.2464 |
| Autumn |   | TSF    | P      | MM-DFS | DWT-LSTM | ARIMA  | SDL    | Winter |   | TSF    | P      | MM-DFS | DWT-LSTM | ARIMA  | SDL    |
| WF1 | RMSE | **4.7464** | 5.5783 | 5.2849 | 5.9684 | 5.9430 | 5.6389 | WF1 | RMSE | **4.7531** | 5.4044 | 5.0058 | 5.1463 | 5.6672 | 5.3935 |
|     | MAE  | **3.5090** | 3.8938 | 3.7003 | 4.3109 | 4.3409 | 4.1522 |     | MAE  | **3.2343** | 3.5594 | 3.3147 | 3.3729 | 3.9695 | 3.8378 |
| WF2 | RMSE | **3.6748** | 4.4854 | 4.0012 | 3.9170 | 4.6135 | 4.4402 | WF2 | RMSE | **4.5551** | 4.8692 | 4.7059 | 4.7371 | 5.1203 | 4.7360 |
|     | MAE  | **2.6277** | 3.2049 | 3.0391 | 2.8903 | 3.3434 | 3.2554 |     | MAE  | **3.1276** | 3.2013 | 3.2246 | 3.2178 | 3.5799 | 3.2875 |
| WF3 | RMSE | **4.3834** | 6.296  | 4.5180 | 4.4948 | 6.7510 | 6.1263 | WF3 | RMSE | **4.6874** | 5.2915 | 5.0463 | 4.9870 | 5.3920 | 5.1397 |
|     | MAE  | **2.8778** | 3.9545 | 2.9372 | 2.9516 | 4.4165 | 4.1650 |     | MAE  | **3.5662** | 3.9521 | 3.8561 | 3.7208 | 4.1393 | 3.9089 |

(The lowest RMSE and MAE values in each case are presented in bold)

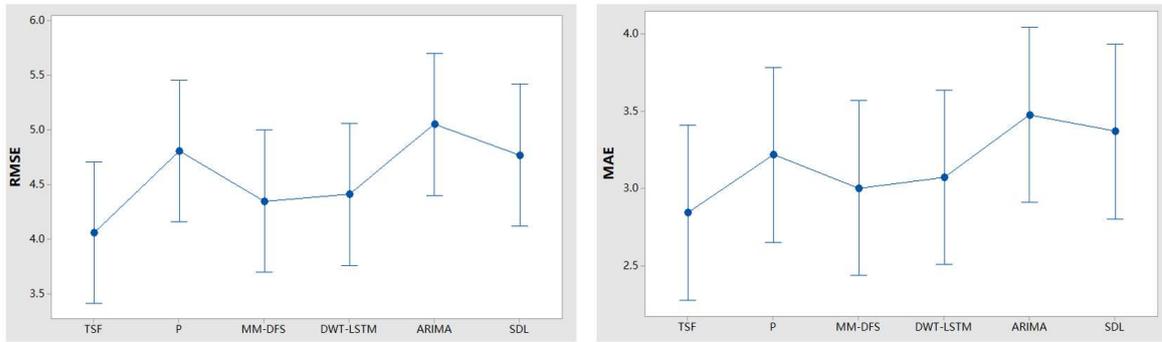

Fig. 12. 95 % confidence interval plot of forecasting accuracies using different forecasting methods.

Table 8. Absolute difference analysis between real wind power and forecasting results.

|          | TSF vs WP | P vs WP | MM-DFS vs WP | DWT-LSTM vs WP | ARIMA vs WP | SDL vs WP |
| -------- | --------- | ------- | ------------ | -------------- | ----------- | --------- |
| Variance | **8.5034** | 13.1778 | 9.3517 | 9.6851 | 13.8744 | 11.7410 |
| Mean     | **2.6487** | 2.9593 | 2.8333 | 2.8049 | 3.2395 | 3.3701 |

(The lowest variance and mean of the absolute difference is presented in bold)

## 5 Conclusions

WPF using an ensemble model is an extensively studied topic; however, several factors have not been considered within this framework. Therefore, in this paper, using the two-stage framework, we first introduced deep neural networks in the first stage to learn data features from different perspectives. Then, we explored the model extrapolation issue that occurred in the second stage, which has been ignored by the WPF community. Finally, we considered a validation set and incorporated the two-stage framework into a moving window based training data updating algorithm, which had two moving window processes to handle the overfitting issue.

Our experiments were divided into three parts. The first experimental results demonstrated that the
17

single-input-multiple-output model obtained forecasting accuracy than the other three models, which indicates that combining historical wind data and NWP data as a single input and adding wind speed as output is likely to improve forecasting accuracy. In addition, the results of the second experiment demonstrated that using SVR and GPR in the second stage might result in model extrapolation errors, which highlights the importance of selecting a suitable ML algorithm to integrate the forecasting results from the first stage. The results of the third experiment demonstrated that the proposed two-stage algorithm generated better and more stable forecasting results than existing methods, which implies the advantages of the proposed algorithm.

The proposed forecasting algorithm could be beneficial to the intraday power market from different perspectives. First, improved forecasts could reduce the balancing costs for power system operators. Second, improved forecasts could benefit wind energy traders in decision-making relative to purchasing costs in a microgrid system. Third, improved forecasts could reduce the penalty costs for wind plant owners because power systems typically require owners to submit short-term WPF in the intraday market. Overall, such an improved model is expected to increase the reliability of power systems and simultaneously offer benefits to energy traders and wind plant owners.

In future, probabilistic WPF based on the proposed TSF algorithm should be investigated. Future work could also focus on reducing the extrapolation issue of other comprehensive algorithms in the second stage.

## Acknowledgments

This study was supported by the Heilongjiang Provincial Postdoctoral Science Foundation (Grant No. LBH-Z19146), the China Postdoctoral Science Foundation (Grant No. 216734), and the National Natural Science Foundation of China (Grant No. 91846301).

## Declaration of Competing Interest

The authors declare that they have no conflicts of interest.

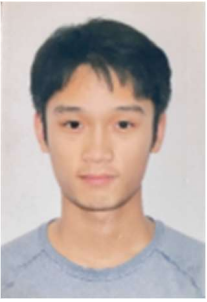

**Jiancheng Qin** is currently pursuing the B. S degree in Computer Science with Harbin Institute of Technology, Heilongjiang, China. His current research interests include deep learning, dynamic programming and reinforcement learning with their applications in the control and management system.

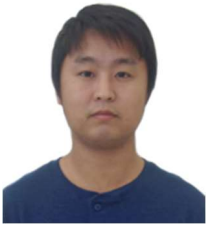

Ying Chen received the Ph.D. degree in Industrial Engineering from University of Texas at Arlington. He is currently an Assistant Professor in the School of Economics and Management at Harbin Institute of Technology. His research interests data mining, machine learning, decision-making under uncertainty and optimization.